\documentclass[12pt]{article}
\pdfoutput=1
\usepackage{natbib}
\usepackage{url}
\usepackage{graphicx}
\begin{document}
\title{GemTools: a fast and efficient approach to estimating genetic ancestry}
\author{Lambertus Klei\,$^{1}$ \and
Brian P. Kent\,$^{2}$ \and
Nadine Melhem\,$^{1}$ \and 
Bernie Devlin\,$^{1}$ 
\and 
Kathryn Roeder\,$^{2,}$\footnote{to whom correspondence should be addressed}
}
\maketitle


\begin{abstract}
  \setcounter{footnote}{1}
\subsection*{Motivation:}
To uncover the genetic basis of \footnotetext{Department of Psychiatry, University of Pittsburgh School of Medicine, Pittsburgh, PA 15213, 
$^{2}$Department of Statistics, Carnegie Mellon University, Pittsburgh, PA 15213}
  complex disease, individuals are often measured at a large number of
  genetic variants (usually SNPs) across the genome.  GemTools
  provides computationally efficient tools for modeling genetic
  ancestry based on SNP genotypes.  The main algorithm creates
  an eigenmap based on genetic similarities, and then clusters subjects
  based on their map position. This process is continued iteratively
  until each cluster is relatively homogeneous. For genetic
  association studies, GemTools matches cases and controls based on
  genetic similarity.  


\subsection*{Availability:}
 GemTools source code, documentation, and
  additional examples
  are available at \url{http://wpicr.wpic.pitt.edu/WPICCompGen/}
\subsection*{Contact:} \url{roeder@stat.cmu.edu}
\end{abstract}

\section{Introduction}
Genetic variants differ in allele frequency by populations, creating
stratification due to differential ancestry. As a consequence,
case-control studies are susceptible to spurious associations between
genetic variants and disease status \citep{lander1994}.  A large panel
of genetic markers can be used to construct an eigenmap, which encodes
the ancestry via the relative genetic similarities within the sample.
Given ancestry coordinates, the effects of population stratification
can be controlled by matching cases and controls \citep{luca2008} or
regressing out ancestry \citep{price2006}.

In {\it GemTools}, eigenmaps are constructed using a spectral clustering
techniques \citep{lee2010}, an approach that is closely related to
principal component analysis \citep{price2006}.  The first computational
challenge involves calculation of the inner product matrix $XX^t$,
where $X$ is an $n\times m$ matrix, indexed by $n$ subjects and $m$
genetic markers.  Next, we require the spectral decomposition of an
$n\times n$ matrix. Often the data include tens of thousands of
subjects and markers making these computations both slow and memory
intensive.  To circumvent these challenges we use a divide and conquer
approach that works by clustering individuals of like ancestry and
then finding eigenmaps for each cluster.  In addition to 
computational efficiency, we find that focusing on fine scale
structure within clusters facilitates matching of cases and controls
of similar ancestry. Thus our approach is faster, more informative, and more accurate
than a brute force computational treatment involving the calculation of
a single eigenmap of the whole dataset \citep{crossett2010}.

\section{Description}

The main function of {\it GemTools}, {\it dacGem}, exploits two
features of the problem.  First, for most applications, the eigenmap
is not of inherent interest; the end goal is to cluster subjects of
like ancestry.  Indeed, if the sample is drawn from subjects of highly
disparate ancestry, an eigenmap of the entire sample will be high
dimensional, making it difficult to match, on a fine-scale, subjects
of similar ancestry \citep{crossett2010}.  Alternatively, if the
sample is recursively partitioned into subsets of similar ancestry,
then a low dimensional local eigenmap can be estimated for each
cluster to determine fine scale structure.  Based on a local eigenmap,
the cases and controls can be can be matched much more reliably within
each cluster.  Second, to produce these clusters in a computationally
efficient way, it is not necessary to compute the eigenmap using the
full sample of $n$ subjects. First we select a {\it base} sample of
$N<n$ representative subjects. Using this fraction of the data, we
construct an eigenmap.  Applying Ward's algorithm, we cluster the base
subjects into subsets with relatively similar ancestry.  Using the
Nystrom approximation \citep{crossett2010} we project the remainder of
the sample onto the map.  Finally, each non-base subject is
incorporated into the cluster of its nearest base neighbor.  This
process is subsequently repeated within each cluster until the
resulting subclusters are small and homogeneous. With this divide and
conquer approach we restrict the most intensive calculations to the
base sample and hence the time and memory required are greatly
reduced. At the same time, as the algorithm progresses to smaller and
smaller clusters, all of the subjects within a cluster can be
considered as part of the base sample without incurring any
substantial computational cost, and the Nystrom approximation is no
longer necessary.

Our approach builds on the {\it Spectral-GEM} algorithm 
\citep{lee2010,crossett2010},
which provides a context for constructing an eigenmap and subsequent
clustering of relatively homogeneous subjects.  This algorithm
determines the number of significant dimensions $D$ required for each
eigenmap. A genetically homogeneous
cluster is defined as one for which the spectral decomposition
includes no significant eigen-values \citep{patterson2006}.

The {\it dacGem} Algorithm proceeds as follows:
\begin{enumerate}
\item Randomly select $N$ subjects and {\it mark} them.
\item Construct a $D-$dimensional eigen-map of the marked subjects.
\item Based on the eigen-map, form homogeneous clusters of subjects.
\item Project unmarked subjects onto the eigen-map.
\item Group unmarked subjects with nearest cluster in the eigen-map.
\end{enumerate}
Repeat steps 1-5 for each cluster consisting of $B$ or more subjects,
until all of the clusters have less than $B$ members.  

The {\it GemTools} functions are implemented in R and can be run in a
variety of operating systems, including Windows, Linux and MacOS. Run
time is linear in the number of markers ($m$) and sub-quadratic in
subjects ($n$). {\it GemTools} provides a computationally efficient alternative to Eigenstrat \citep{patterson2006}.

\section{Example}

Although designed for large data sets, to illustrate {\it GemTools} in
detail we used a small sample of publicly available data from the
Human Genome Diversity Project \citep{rosenberg2002}.  Specifically,
we focus on 226 individuals from two continents (Africa and Europe)
with 4 and 7 ethnicities representing each continent, respectively. The
African ethnicities are Biaka Pygmies, Mandenka, Mbuti Pygmies, and
Yorubans.  Ethnicities representing Europe are Adygei, French, French
Basques, Orcadians, Italians, Russians, and Sardinians. From the
genome wide platform, and for illustrative purposes, we selected only
1,167 nearly independent SNPs with minor allele frequency greater than
5\% (data posted with software).  Figure 1 illustrates the outcome of
{\it dacGem} using a base sample of $N=100$ and a maximum cluster size
of $B=50$.  The first split separates subjects by continental
ancestry.  The second and third splits separate many subjects
correctly by ethnicity.

\begin{figure}[!tpb]
\centerline{\includegraphics[trim = 10mm 30mm 0mm 25mm,clip]{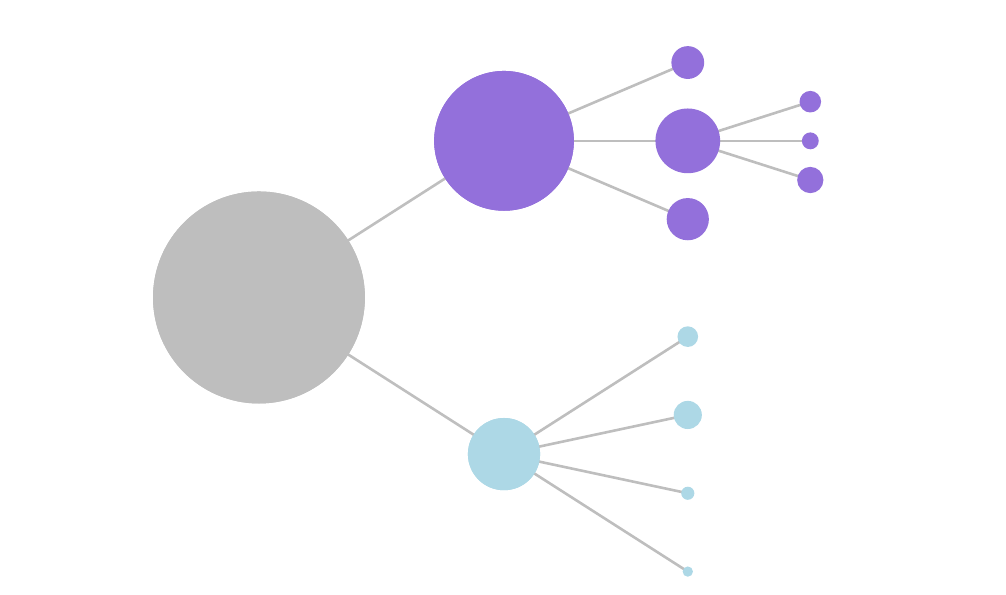}}
\caption{Iterative clustering using dacGem on the Human
Genome Diversity Project data.  The size of each ball is proportional
to the number of subjects in the cluster.  In the first split, using 1
significant dimension ($D=1$), the sample is split into 2 clusters by
continent.  Using $D=2$ dimensions the European cluster (purple) is
split into 3 clusters (primarily Sardinians, French Basques, and all
others).  In the next stage, the mixed group is further split into 3
clusters (primarily Adygei, Russian and a mix of the remaining
groups). Using $D=3$ dimensions the African cluster (blue) is split
into 4 clusters conforming fairly closely to the 4 ethnicities.}\label{fig:01}
\end{figure}

We also applied this method to a multi-ethnic sample including
approximately 20,000 subjects with $12,000$ markers, and obtained good
results. Using a base sample of $N=500$ and maximum cluster size
of $B=1000$ the run time for this large dataset was 45 minutes with
memory use of 6 Gb.  Direct eigen-analysis and hierarchical clustering
of a matrix of order 20,000 is impractical.  In its standard
implementation, GemTools never uses a matrix of order greater than
1,000 for these two operations.  In most steps of the algorithm,
matrices are of order less than 500.

Once a cluster with $<B$ members has been identified, there are two
choices for the final step of genetic association analysis.  Using the
{\it ccMatchGem} function an eigenmap for each cluster is calculated.
Cases and controls of like ancestry are matched.  The {\it ccMatchGem}
algorithm matches subjects using a $d$-dimensional eigenmap, where $d
= \max(D,D^*)$, and $D^*$ is a user-specified minimum threshold.
Alternatively, the clustering process can be continued recursively
using the function {\it clusterGem} until each cluster has $D=0$
significant dimensions. In this scenario, one can consider all of the
subjects within a sub-cluster as being from a common ancestry and use
cluster membership as a factor variable for stratified analysis of the
data.

\section*{Acknowledgement}
\paragraph{Funding:} This work was funded by the National Institutes of Health grant MH057881.
Data were obtained from the HGDP high-resolution genome-wide SNP panel.


\end{document}